\documentclass[aps,pra,twocolumn,floatfix,longbibliography, superscriptaddress]{revtex4-1}

\usepackage{pifont}
\usepackage[T1]{fontenc}
\usepackage[dvipsnames]{xcolor}
\definecolor{bananayellow}{rgb}{1.0, 0.88, 0.21}
\definecolor{amethyst}{rgb}{0.6, 0.4, 0.8}
\definecolor{ao(english)}{rgb}{0.0, 0.5, 0.0}

\usepackage{float}
\usepackage{epsfig}
\usepackage{bm}
\usepackage[matrix,frame,arrow]{xy}

\usepackage[applemac]{inputenc}
\usepackage[T1]{fontenc}
\usepackage{lmodern}
\usepackage[english]{babel}
\usepackage{amsmath}
\usepackage{ae}
\usepackage{units}
\usepackage{amssymb}
\usepackage{color}
\usepackage{graphicx}
\usepackage{bbm}
\usepackage{url}
\usepackage{nomencl}
\usepackage{subfigure}
\usepackage{slashed}
\usepackage{soul}

\usepackage{tikz}
\usetikzlibrary{arrows,snakes,calc}

\usepackage{fixmath}
\usepackage{booktabs}
\usepackage{mathtools}

\newcommand{\bra}[1]{\langle #1|}
\newcommand{\ket}[1]{|#1\rangle}

\newcommand{\figref}[1]{\mbox{Fig.~\ref{#1}}}

\newcommand{\secref}[1]{\mbox{Sec.~\ref{#1}}}

\newcommand{\appref}[1]{\mbox{Appendix~\ref{#1}}}
\renewcommand{\eqref}[1]{\mbox{Eq.~(\ref{#1})}}

\newcommand{\be}{\begin{equation}}
\newcommand{\ee}{\end{equation}}
\newcommand{\bea}{\begin{eqnarray}}
\newcommand{\eea}{\end{eqnarray}}

\newcommand{\Ak}{\mathcal{A}_\kappa}

\newcommand{\Gg}{\mathcal{G}_\gamma}
\newcommand{\Gk}{\mathcal{G}_\kappa}

\usepackage[colorlinks]{hyperref}
\hypersetup{%
	plainpages=true,
	breaklinks=true,
	hypertexnames=false,
	pageanchor=true,
	colorlinks=true,
	linkcolor={blue},
	citecolor={red},
	urlcolor={blue},
	anchorcolor={black}
}

\begin{document}

\title{Spin squeezing by one-photon-two-atom excitations processes in atomic ensembles}

\author{Vincenzo Macr\`{i}}
\email[e-mail:]{vincenzo.macri@riken.jp}
\affiliation{Theoretical Quantum Physics Laboratory, RIKEN, Wako-shi, Saitama 351-0198, Japan}
\author{Franco Nori}
\affiliation{Theoretical Quantum Physics Laboratory, RIKEN, Wako-shi, Saitama 351-0198, Japan}
\affiliation{Physics Department, The University of Michigan, Ann Arbor, Michigan 48109-1040, USA}
\author{Salvatore Savasta}
\affiliation{Theoretical Quantum Physics Laboratory, RIKEN, Wako-shi, Saitama 351-0198, Japan}
\affiliation{Dipartimento di Scienze Matematiche e Informatiche, Scienze Fisiche e  Scienze della Terra, Universit\`{a} di Messina, I-98166 Messina, Italy}
\author{David Zueco}
\affiliation {Instituto de Ciencia de Materiales de Arag\`{o}n and Departamento de F\`{i}sica de la Materia Condensada, CSIC-Universidad de Zaragoza, 50009 Zaragoza, pain}
\affiliation{Fundaci\`{o}n ARAID, Campus R\`{i}o Ebro, 50018 Zaragoza, Spain}

\date{\today}

\begin{abstract}
It has been shown elsewhere that two spatially separated atoms can jointly absorb one photon, whose frequency is equal to the sum of the transition frequencies of the two atoms. 
We describe this process in the presence of an ensemble of many two-level atoms, and show that it can be used to generate spin squeezing and entanglement.
This resonant collective process allows to create a sizeable squeezing already at the single-photon limit. 
It represents a novel way for generating many-body spin-spin interactions, yielding  a two-axis twisting-like interaction among the spins, which is very efficient for the generation of spin squeezing.
We perform explicit calculations for ensembles of magnetic molecules coupled to a superconducting coplanar cavities. 
This system represents an attractive on-chip architecture for the realization of improved sensing.
\end{abstract}


\maketitle


\section{Introduction}

Recently, it has been shown how a large number of nonlinear optics processes can be realized with individual two-level atoms coupled to one or more resonator modes in the ultrastrong coupling regime (USC)\cite{Kockum2017}, where the coupling strength starts to become comparable to the resonance
frequencies of the bare system components. In this regime, counter-rotating terms in the light-matter interaction Hamiltonian start to play a role and enable novel higher-order
processes \cite{kockum2019,Forn2019}.
These  vacuum-boosted nonlinear-optics implementations, in contrast
to conventional realizations of various multi-wave
mixing processes in nonlinear optics, can reach perfect
efficiency, need only a minimal number of photons, and
require only two atomic levels.

Many of these processes can be described
in terms of higher-order perturbation theory, in which the system passes from an
initial state $| i \rangle$  to the final state $| f \rangle$  (with the same energy), via a number of virtual
transitions to intermediate states. When the light–matter
coupling strength increases, the vacuum fluctuations of
the electromagnetic field become able to induce efficiently such virtual
transitions, replacing the role of the intense applied
fields in conventional nonlinear optics. In this way, higher-order
processes involving counter-rotating terms can create
an effective coupling between two states of the system
($| i \rangle$ and $| f \rangle$) which can have different number of excitations.
The strength of the resulting effective coupling scales approximately as $g_{\rm eff} \sim g (g/\omega)^n$, where $g$ is the light-matter coupling strength, $\omega$ describes a bare resonance frequency of the light or matter component, and $n$ is the number of involved virtual transitions.
However, the required light-matter coupling strength to observe these deterministic nonlinear optics effects with the minimum amount of photons, can only be reached, at least presently, only with superconducting artificial atoms.
Recently, simulations of these nonlinear optics effects with strong-coupling systems (with $g/\omega\ll1$) dressed by classical drives have been proposed \cite{Ballester2007,Pedernales2015,braumuller2017,Dingshun2018,munoz2019}.

Here, we propose a different route, already widely explored to observe linear optical processes in the USC regime: the light–matter coupling strength can be enhanced by increasing the
number $N$ of emitters interacting with the resonator. The resulting collective coupling strength scales as $g \sqrt{N}$. In this way the USC regime can be reached in a wider range of systems (see, e.g., \cite{Schwartz2011,Kena-Cohen2013,gubbin2014low,mazzeo2014ultrastrong,Gambino2014,Anappara2009,genco2018bright,Gunter2009,Todorov2010,Scalari2012,Maissen2014,zhang2016collective,li2018vacuum}).

One of the most interesting nonlinear optical effects predicted in the USC regime consists of the simultaneously excitation of two or more spatially separated atoms by a single photon \cite{Garziano2016,Zhao2017,Nori2017}.
This process is reversible, so that the atoms can return to a lower-energy state by collectively emitting one photon.
This is a two-atom resonant process occurring when the atom transition is half of the cavity frequency.
Here, \emph {we generalize this process to many two-level atoms} which can be described as an ensemble of $N$ pseudo-spins.. 
This opens the way to investigate cases in which several photons can excite different atomic pairs, producing an effective, controllable interaction among spins.
As we will show, the simultaneous excitation of spin pairs in a large ensemble, can give rise to multi-atom entanglement and to strong spin squeezing, which could be useful for application in quantum technologies.

Quantum sensors beat the shot-noise limit of precision by using entangled states.
%
%
Pseudo spin-$1/2$ ensembles, serving as probes for measuring a magnetic field, represent a paradigmatic example \cite{Toth2014}.
If they are prepared in a  convenient squeezed-entangled state, the minimized quadrature reduces the precession angle uncertainty and, thus, the error in the field estimation \cite{Ma2011}.
The Heisenberg  principle imposes the ultimate scaling error as  $1/N$, with $N$ the number of spins.
Therefore,   preparing a macroscopic spin state in a highly squeezed state is {\em a key resource}  for  quantum metrology.

Squeezed states in atomic ensembles are prepared by inducing interactions among the spins.  Different schemes, including atom-atom interactions in traps, feedback and projective measurements have shown up to -18 dB quadrature reduction 
\cite{riedel2010, Leroux2010, Wasilewski2010, Muessel2014, SchleierSmith2010, Bohnet2014, Vasilakis2015, Hosten2016a}.

An interesting alternative is the \emph {deterministic} production  within one-axis twisting  interactions generated inside a cavity  \cite{Hosten2016, Norcia2018}. So far, the reported results are limited to  -8 dB.
Thus, it is desirable to find improved but  \emph{deterministic} protocols, as the preparation of  initial coherent states superpositions \cite{LewisSwan2018}.

Another alternative involves considering two-axis twisting Hamiltonians  that are known to be optimal for generating squeezing \cite{Opatrn2015, Liu2011}. However, it remains to show their advantage in presence of noise and decoherence \cite{Borregaard2017}. It has been shown that collective decoherence can be suppressed via continuous dynamical decoupling \cite{Chaudhry2012}. This approach can make spin squeezing more robust to noise and closer to the Heisenberg limit (optimal squeezing, scaling a  $\xi^2 \sim 1/N$).

In this work we show that the one-photon-two-atom excitations process can determine approximate two-axis twisting interactions in  spin ensembles coupled to a cavity mode, which are known to generate optimally squeezed states. 

Our protocol is different from previous approaches in several ways. Although the light-matter system is in the dispersive regime, it is a  resonant mechanism involving \emph{real} photons. This is complementary to the case where the field can be integrated out, so that the spin-spin interactions are mediated by virtual photons \cite{Agarwal1997, Srensen2002, Borregaard2017}.
The approach proposed here is  a third-order nonlinear optical process boosted by virtual photons \cite{Kockum2017}. Therefore, for obtaining a significant rate for the excitation of atom pairs, even for weak input fields, the light-atoms coupling rate should reach a significant fraction of the atomic transition frequency.

However, in this work, the relevant coupling strength is not the single-spin coupling, $g$, but the effective \emph{collective interaction} between the $N$ spins and the single-cavity mode, which yields the scaling $ \sim g^3 N $. This widely broadens the number of platforms where this process could be implemented.

Moreover, it is also required that the atomic or molecular potential does not display inversion symmetry. In this case, the system can be described by an extended Dicke model, with atoms displaying both longitudinal and transverse coupling with the cavity mode (see, e.g., Ref.s~\cite{Garziano2015,Garziano2016,Lambert2018}).

Finally, real photons must be injected inside the cavity. The drive could be even a coherent resonant field. Unexpectedly, we find that even a single cavity photon is able to generate a significant amount of squeezing. A final advantage of using resonant real photons is that the interaction can be controlled by acting either on the driving or on the atoms-cavity detuning.

Our results could be implemented using several kinds of qubits as, e.g., cold atoms, chiral molecules or superconducting flux qubits.  One interesting architecture consists of hybrid spin-superconductor systems \cite{Xiang2013a}, where the spins can be NV-centers \cite{Schuster2010, Kubo2010, Wu2010, Amsss2011,XinYou2013,Xiang2013} or  more general spins \cite{Jenkins2013, jenkins2016} that are coupled to a superconducting microwave resonator.
%

\begin{figure}[!ht]
	\centering
	\includegraphics[width=0.9\columnwidth]{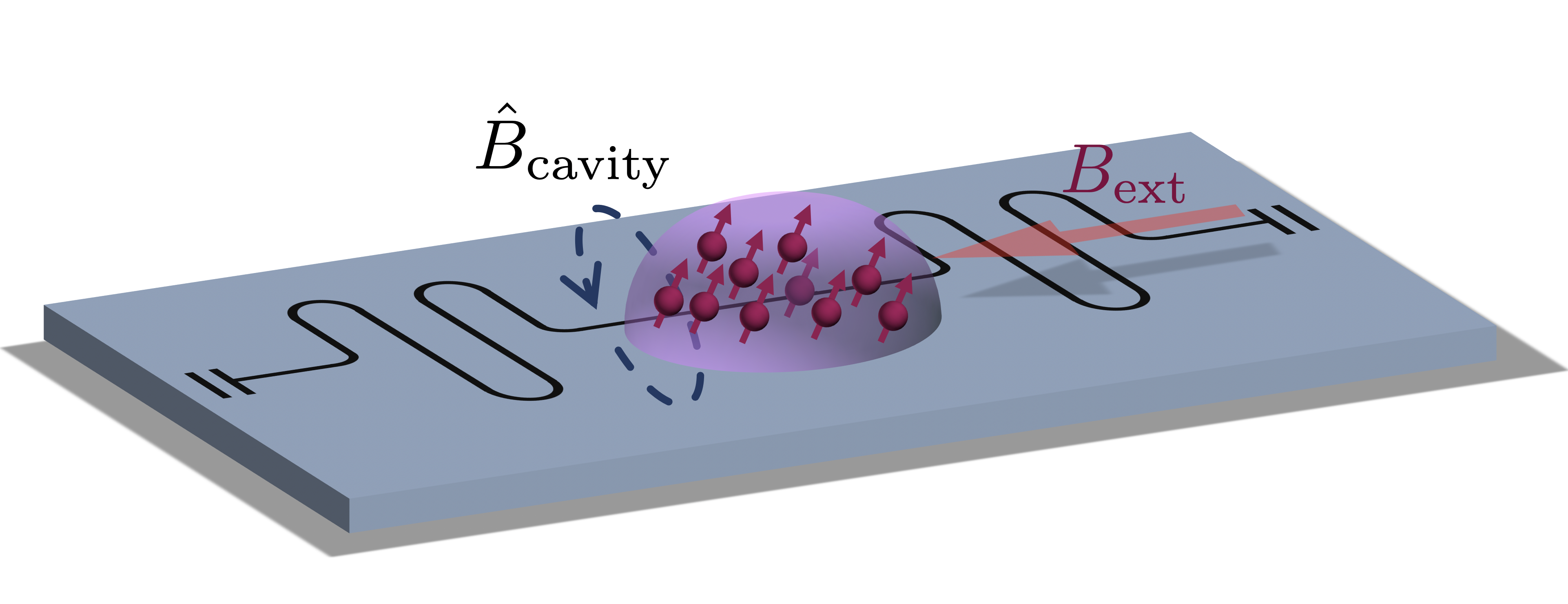} 
	\caption{\footnotesize
		Sketch for the proposed architecture.  An ensemble of magnetic molecules (red in the figure) are deposited on top of a  superconducting coplanar waveguide resonator.  The spin-cavity coupling is mediated by the magnetic field generated by the stationary currents in the circuit, denoted in the figure as $\hat B_{\rm cavity}$.  The spins transition frequency can be tuned by means of external control fields, $\hat B_{\rm ext}$.}
	\label{fig:setup}
\end{figure}
\section{Resonant excitation of atomic pairs: effective Hamiltonian and squeezing parameter}
\begin{figure}[!t]
	\centering
	\includegraphics[width=1\columnwidth]{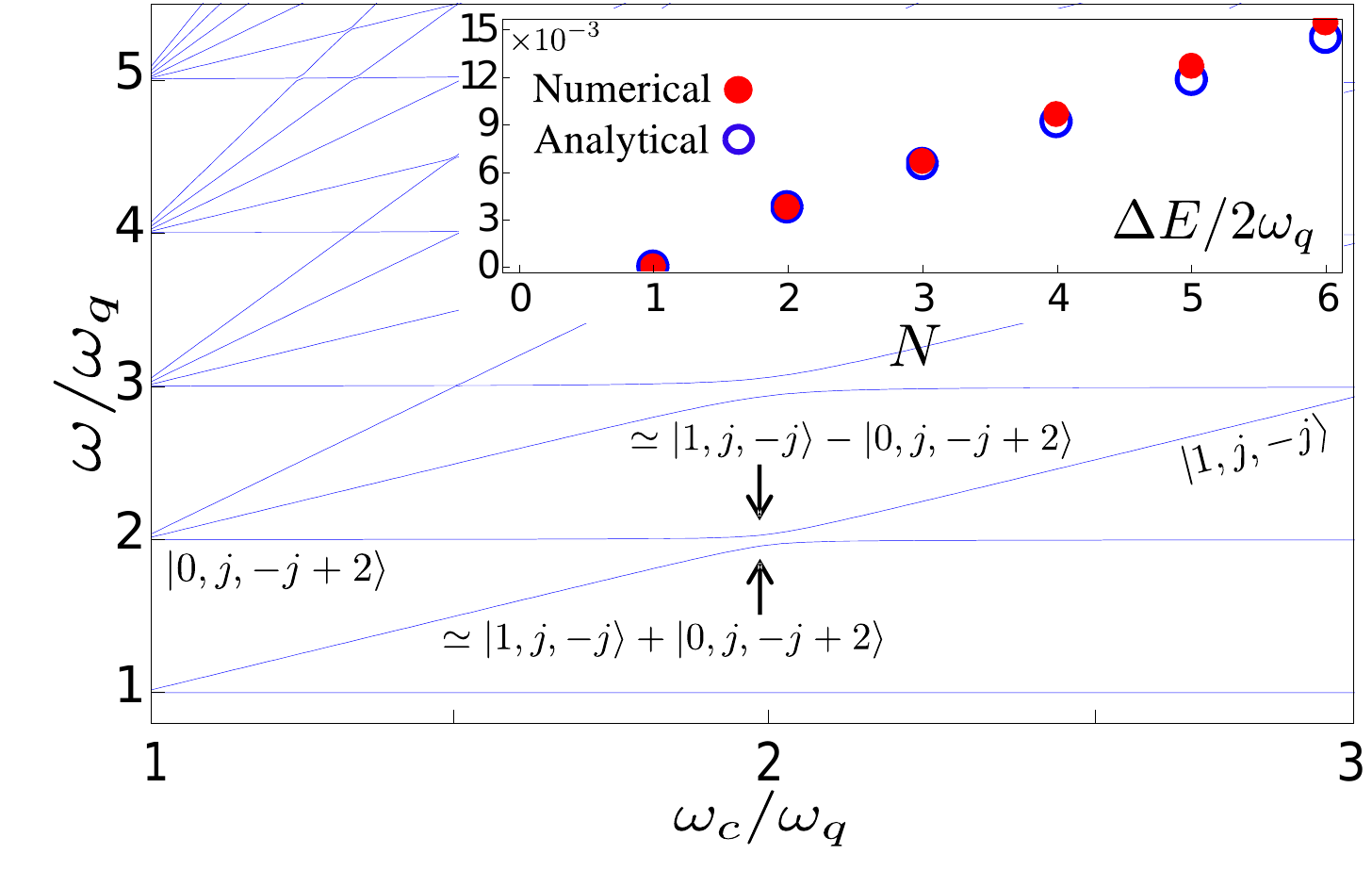}
	\caption{\footnotesize Lowest energy levels of the system Hamiltonian obtained for  $N=20$ qubits as function of  the ratio between the cavity frequency $\omega_c$ and that of identical qubits $\omega_q$. We used  $ g = 2.5\times 10^{-4}\,\omega_q $ and $\theta= \pi/6$. The inset compares the analytical (blue dots) and numerical (red dots) results for the effective coupling between the states $ |0,j,-j+2\rangle  $ and $ |1,j,-j\rangle  $ versus the number of qubits.
	}
	\label{fig:spectrum}
\end{figure}

We consider an ensemble of $N$ identical two-level systems  equally coupled to a single-mode cavity. The Hamiltonian can be written as:
\be
\label{H}
\hat H = \Delta \hat J_z +\epsilon \hat J_x
+ \omega_{\rm c}\, \hat a^\dag \hat a +2 g (\hat a + \hat a^\dagger) \hat J_x\, ,
\ee  
where $\hat a$ and $\hat a^\dag$ are the usual operators for cavity photons, while $2\hat J_\alpha =\sum_i \hat \sigma_\alpha^{i}  $ ($\alpha=x,y,z$) are the collective angular momentum operators. As a consequence of the latter expression, lowering and raising spin operators are defined as $\hat J_\pm = \sum_ i \hat \sigma^i_\pm$.
Parity symmetry breaking is described by the second term in \eqref{H}. For $\epsilon= 0$ the celebrated Dicke model is recovered \cite{garraway2011,Shammah2017,Shammah2018,kirton2018}.
When $\epsilon \neq 0$, $\hat H$ can couple states differing by an odd number of excitations. For example, an avoided level crossing, originating from the coupling of the states  $\hat a^\dagger|0, j, -j \rangle \leftrightarrow \hat J_+^2 |0, j, -j \rangle$, is expected  when the resonance frequency of the cavity $\omega_c \simeq 2\omega_q = \sqrt{\Delta^2 + \epsilon^2}$.
We label the states as  $ | n, j, m \rangle $, where the quantum number $n$ describes the Fock states of the cavity, and $j= N/2$ is the total angular momentum and $m = -j + N_{\rm exc}$ is the $\hat J_z$ eigenstate, where $N_{\rm exc}$ describes the number of excited atoms.
Such a coupling has been described for the two qubit case ($N=2$) only in \cite{Garziano2016}.  
Figure~\ref{fig:spectrum} confirms that it also occurs for any $N \geq 2$.  
The analysis of those processes non-conserving the number of excitations can be simplified, deriving an effective Hamiltonian by using perturbation theory \cite{Shao2017}. 
For frequencies close to the resonance condition $\omega_c \simeq 2 \omega_q$, from  \eqref{H}, as shown in Appendix \ref{app:effective}, the following effective interaction Hamiltonian can be obtained:
\be \label{Heff} 
\hat H_{\rm eff} = g_{\rm eff}
\left(\hat a\hat J_{+}^{2}+\hat a^{\dag}\hat J_{-}^{2}\right)\, , 
\ee
where
\be \label{geff}
g_{\rm eff} =
-\frac{4g^3 \cos^2 \theta \sin \theta}{3 \omega^2_{ q}}\, ,
\ee
with $\sin \theta= \epsilon /\sqrt{\Delta^2 + \epsilon^2} $. This procedure also gives rise to a renormalization of the atomic frequencies, which can be reabsorbed into $\omega_q$.

The effective Hamiltonian \eqref{Heff} yields a series of nonzero transition matrix elements $\langle n-1, j,m+2 | \hat H_{\rm eff}| n, j, m \rangle$, determining a ladder of avoided level crossings at $\omega_c =2 \omega_q$, with energy splittings which are twice these matrix elements. The lowest energy splitting is 
\be
\Delta E =  2 g_{\rm eff} \sqrt{2N(N-1)}\,.
\ee
It is worth noticing that, for large $N$, this splitting scales as $ \sim\! g^3 N$. 
%
The comparison between $\Delta E/ 2$ and the corresponding half-splitting energy obtained from the exact numerical diagonalization of the Hamiltonian in \eqref{H}  is shown in the inset in \figref{fig:spectrum}.  
We note the very good agreement.  
%
Figure~\ref{fig:spectrum} displays the lowest excited energy levels of the effective Hamiltonian (the ground state energy is zero) as a function of $\omega_c/ \omega_q$, obtained for $N=20$ qubits. Avoided level crossings are clearly visible at $\omega_c  \simeq 2 \omega_q$.
In \appref{app:Full}, we also compare the higher energy avoided level crossings obtained by using the full \eqref{H} and the effective \eqref{Heff} models. The results still show a quite good agreement for $\omega_c \simeq 2 \omega_q$.

To characterize the squeezing we use the spin-squeezing parameter $ \xi^2 $, proposed by Wineland \emph{et} \emph{al.} \cite{Ma2011,Wineland1992,Wineland1994}, 
\be \label{ss}
\xi^2 = N \frac{\langle ( \hat {\bf J}\cdot  {\bf n}_{\perp})^2 \rangle - \langle  \hat {\bf J} \cdot {\bf n}_{\perp} \rangle^2}{\langle \hat {\bf J} \rangle ^2}\, ,
\ee 
where the unit vector ${\bf n}_{\perp}$ is chosen to minimize the numerator.
This is the ratio between the fluctuations of a general state versus the coherent spin state (CSS), for the determination of the resonance frequency in Ramsey spectroscopy. The CSS here acts as a noise-reference state. For $\xi < 1$ a gain in interferometric precision is possible compared to using a
coherent spin state \cite{Ma2011,Wineland1994}.

\subsection{Single photon squeezing}\label{single_photon}
The simplest and, at the same time, weirdest illustration of the role of  \emph{real} cavity photons in spin squeezing generation  can be obtained by looking at the squeezing generated by a single photon.  
Let us consider as initial state  a superposition of zero and one photon, with all the atoms in their ground state:
$ |\psi (0) \rangle = \cos \varphi  |0,j,-j\rangle+ \sin \varphi  |1,j,-j\rangle $.
Using the effective Hamiltonian in \eqref{Heff} and neglecting the losses, the time evolution can be analytically calculated:
\bea |\psi (t) \rangle &=& \cos \varphi  |0,j,-j\rangle + \sin \varphi  \cos (g_{\rm eff} t)|1,j,-j\rangle \nonumber \\ &-& i \sin \varphi  \sin(g_{\rm eff} t)|0,j,-j+2 \rangle\, ,
\eea
from which the spin squeezing parameter in \eqref{ss} can be obtained. 
We chose $ {\bf n}_{\perp} = (\cos \phi, \sin \phi, 0)$ (orthogonal to the $z$ axis).
In \figref{fig:spin}(a) we plot the time evolution of $	\xi^2$.
We also show the mean photon number
\be
\langle \hat a^{\dag} \hat a\rangle = \sin^2\!\varphi\,  \cos^2 (g_{\rm eff}t)\, ,
\ee
and the mean number of excited atoms
\be
N_{\rm exc} =j+\langle \hat J_z \rangle=2\sin^2 \varphi\, \sin^2 (g_{\rm eff}t)\,.
\ee
We considered a system of $N=20$ spins with parameters $ g=0.115 \,\omega_q $, $\theta = \pi /6$ (i.e., $g_{\rm eff}= 0.01$), and $ \varphi = 0.45\pi$.
We also used the phase $ \phi=\pi/4$, providing the maximum squeezing (corresponding to the minimum value of $\xi^2$).  
When the cavity excitation is completely transferred to the ensemble of two-level systems, the squeezing  reaches its maximum. 
This is because the system is in a  quantum superposition of  the states $|j, -j\rangle$ and $|j, -j+2\rangle$.  
They are the two first components of an even superposition of coherent states, i.e. an {\em entangled} cat state.
Let us emphasize the  maximum amount of quantum-noise reduction obtained,  $  \xi^2  \simeq 0.55$, already with a single photon. Very similar results can be obtained using the full original Hamiltonian \eqref{H}, as reported in \appref{app:Full}.

Notice, that the considered coupling strength $g$ between the resonator and each qubit can be realized experimentally with circuit-QED systems in the USC regime (see, e.g., \cite{Niemczyk2010}. However, since the effective coupling scales linearly with $N$, increasing the number of superconducting qubits, even lower individual coupling strength can become sufficient. Notice that, the coherent coupling of superconducting resonators with very large ensembles of flux qubits have been demonstrated \cite{Kakuyanagi2016}. Moreover, these systems can display decay rates much lower than the resulting effective coupling $\sim g_{\rm eff} {N}$. When the loss rates are equal or larger than five times $\sim g_{\rm eff} {N}$, they do not affect the dynamics shown in \figref{fig:spin}(a).

\subsection{Dissipation and drivings beyond one-photon}
We now move to the scenario where both the spins and the cavity  are affected by dissipation. Assuming that each atom has a $\gamma$ decay channel, and applying the typical second-order Born and Markov approximations, we end up with  the master equation  for the density matrix of the cavity plus spins system [see, e.g., \cite{Gelhausen2017} and Appendix \ref{app:dissipation}]:
\begin{equation}
\label{qme}
\dot {\hat \varrho} = -i [\hat H, \hat \varrho] + \kappa {\mathcal D} [\hat a]  + \frac{\gamma}{N} {\mathcal D}[\hat J_-]\, ,
\end{equation}
where ${\mathcal D} [\hat O] = \hat O \hat \varrho \hat O^\dagger - 1/2 \{ \hat O^\dagger \hat O, \hat \varrho \}$ are the dissipators in Lindblad form, and $\kappa$ and $\gamma$ are the loss rates for the cavity and the spins, respectively. Also, we discuss other kinds of drivings beyond the single-photon example of the preceding section.

\begin{figure}[!ht]
	\centering
	\includegraphics[width=1.05\columnwidth]{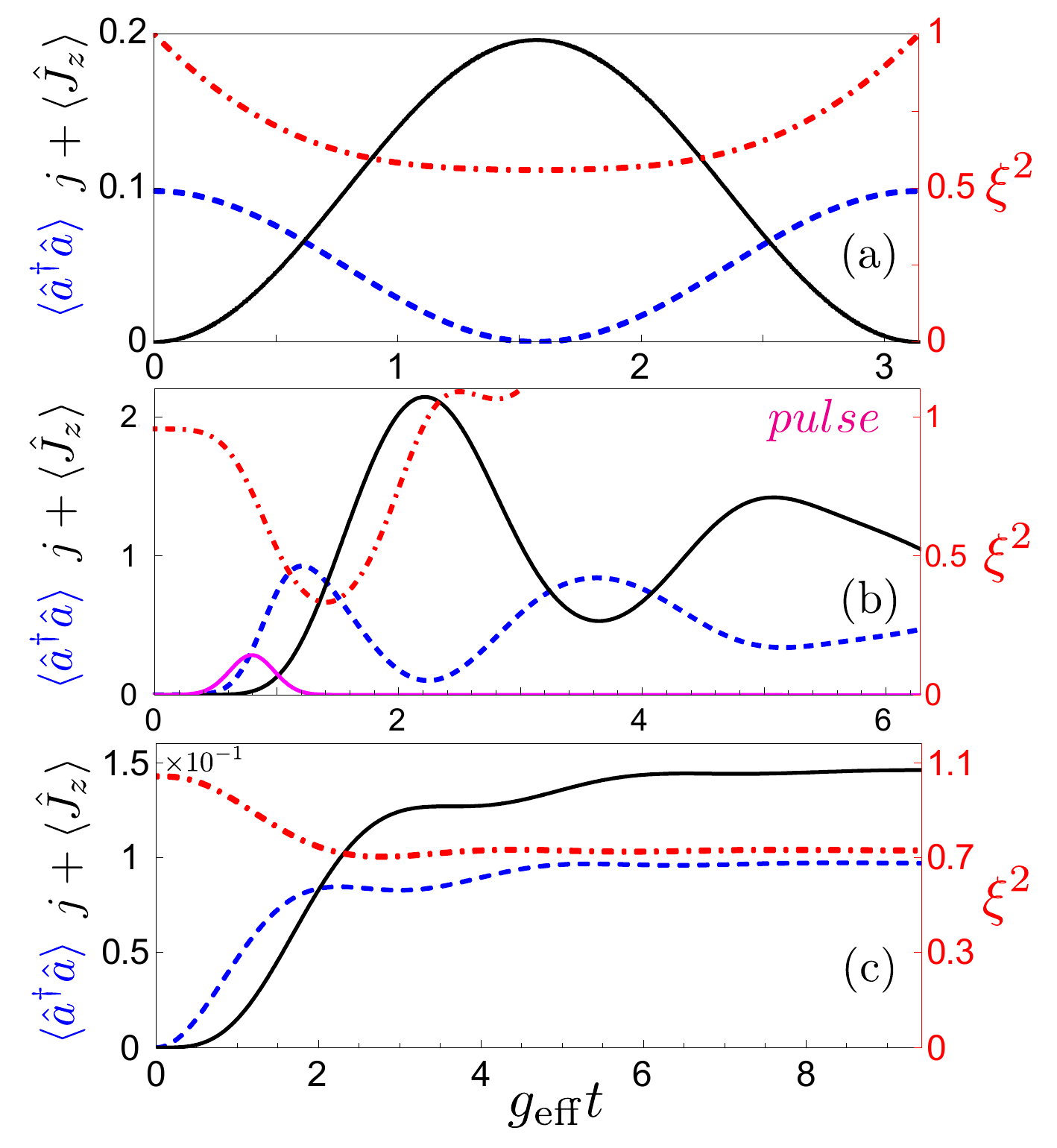}
	\caption{\footnotesize System dynamics for $ \omega_c= 2\omega_q $, (a) single-photon dynamics, (b) subject to a $ \pi $-pulse (magenta solid curve) drive of the cavity mode and (c) under continuous-wave drive of the cavity mode. The blue dashed curve describes the mean photon number $ \langle \hat a^{\dag} \hat a \rangle $, while the black solid curve describes the mean collective spin excitation $ j+\langle \hat J_z\rangle $. The squeezing parameter $ \xi^2 $ is also plotted as a red dot-dashed curve with values given on the $ y $ axis on the right. All numerical parameters are given in the text.
	}
	\label{fig:spin}
\end{figure}

Assuming that the system starts in its ground state,  we first consider a resonant optical pulse feeding the cavity, including an additional time dependent Hamiltonian term $\hat V_d = {\cal F} (t) (\hat a + \hat a^\dag)$, where
$ {\cal F} = {\cal A}\, {\cal G}(t) \cos{(\omega_{\rm d}\, t)}$, with ${\cal G}(t)$ being a normalized Gaussian function.
We consider pulses with their central frequency resonant with the cavity ($\omega_{\rm d} = \omega_{\rm c}$).
Figure~\ref{fig:spin}(b) displays the system dynamics after the pulse arrival, for a system consisting of  a cavity mode and {$N=10$ spins.
	The parameters used are  $ g=0.115\, \omega_q $, $  \gamma=10^{-4} \omega_q $, $ \kappa=\gamma/2 $, $ {\cal A}/ \omega_q= 3 \pi/4 $ and $ \theta= \pi/6 $}.
The Fig.~\ref{fig:spin}(b) shows that the mean photon number $\langle \hat  a^{\dag} \hat a \rangle$, as expected,  is anticorrelated to the mean collective spin excitation $j+ \langle \hat J_z \rangle $.  Here, the even spin states superposition  involves higher spin states of the type $|j -j + 2n \rangle$, which allows to generate a higher degree of spin squeezing. Notice that the obtained maximum spin squeezing (corresponding to the lowest value of the spin parameter $\xi^2$) is quite high, despite the cavity being fed with a weak coherent pulse producing a peak mean intracavity photon number slightly less than one.

We also investigate the case of weak continuous driving:  ${\cal F} = {\cal A}\, \cos{(\omega_{\rm d}\, t)}$  with  ${\cal A}= 2.5\;\gamma $. The other parameters used are, $ g=0.115\, \omega_q $, $  \gamma=10^{-3} \omega_q $ and $ \kappa=\gamma $. In \figref{fig:spin}(c)  we see that squeezing starts to build up when the spin population grows, as expected. In this case, because of the continuous character of the driving, both the cavity and the spins populate to reach a stationary value. The squeezing achieved in this full quantum simulation is quite low, owing to the losses and the low-excitation amplitude.
%

\subsection{Macroscopic spin ensemble}	
\label{MSE}

The full quantum numerical simulations are limited to few tens of qubits.  
However, for most practical purposes $N \gg 1$  is needed. When $N$ is sufficiently large, the collective spin operators can be replaced by a  bosonic mode ($\hat J_+ \to \sqrt{N}\, \hat b$), so that the squeezing parameter defined in \eqref{ss} tends to the bosonic squeezing measure \cite{Ma2011, Wang2003}, 
\begin{equation}
\label{sb}
\xi_{N\to \infty}^2 = 1 + 2(\langle \hat b^\dagger \hat b \rangle - |\langle \hat b^2\rangle|) \, .
\end{equation} 
%
In this limit, the effective Hamiltonian [Cf. \eqref{Heff}] becomes
\begin{equation}
\label{Heffbos}
\hat  H_{\rm eff} = g_{\rm eff} N (\hat a {\hat b}^{\dagger 2} + \hat a^\dagger {\hat b}^2)\, ,
\end{equation}
while the Lindbladians are replaced consistently:~ $\frac{1}{N} {\mathcal D} [\hat J_-] \to {\mathcal D} [\hat b]$ [see \eqref{qme}].
The Hamiltonian in \eqref{Heffbos}, resulting from the bosonization of the effective Hamiltonian in \eqref{Heff}, corresponds to the model Hamiltonian used to describe, in quantum optics, degenerate parametric processes, like second-harmonic generation and parametric down-conversion \cite{Mandel1995,Roy2016}. Specifically, the term $\hat a {\hat b}^{\dagger 2}$ can describe the {\em fission} of a photon of a mode at frequency $\omega_a$ into a photon pair at frequency $\omega_b$ such that $\omega_a = 2 \omega_b$; while the term $\hat a^\dagger {\hat b}^2$ describes the opposite process, where two photons at frequency $\omega_b$ converts into a photon of the mode at frequency $\omega_a = 2 \omega_b$. Notice that the Hamiltonian in \eqref{Heffbos} also applies to the description of parametric processes in completely different systems (see, e.g., \cite{Macri2018}).

We observe that the resulting coupling in \eqref{Heffbos} scales as $g^3 N$. Such bosonic approximation is expected to work fine in the limit $g \to 0$ and $N \to \infty$, such that  $g^3 N \to$ constant. However, we point out that this approximation can work well even in the more physical case of a finite number of atoms, when the number of excitations in the system are significantly smaller than the number $N$ of atoms in the ensemble.
Notice that, such limit is different from the usually considered thermodynamic limit of the Dicke model (see, e.g. \cite{Emary2003}), where $g \sqrt{N} \to$ constant is assumed.
If the latter would have  been assumed, the resulting coupling strength $g_{\rm eff} N$ in \eqref{Heffbos} would go to zero. This is not a surprise, since it is known that in the limit $g \to 0$ and $N \to \infty$, and $g \sqrt{N} \to$ constant, optical nonlinearities in the Dicke model disappear. 

We also observe that the Dicke model in the dispersive regime gives rise to energy shifts of both the atomic and cavity resonances scaling as $g^2N$, which we limit to take into account phenomenologically, just adjusting the bare energy levels. Hence, increasing the number of atoms, while keeping constant the resulting coupling strength $g_{\rm eff} N$, determines relevant energy shifts which have to be taken into account.

In order to maximize squeezing, a strong driving is required. In this case, a direct simulation,
using the full quantum models in \eqref{H}, \eqref{Heff}, or also  in \eqref{Heffbos} becomes not feasible. Hence, we start from the dynamics induced by the Hamiltonian in \eqref{Heffbos} and apply the mean-field approximation  in order to describe the cavity-spins dynamics \cite{Roy2016}. 

In experiments, it is possible to freeze the squeezed state  at the time when the maximum squeezing is reached, by detuning the spins transition frequency. 
Then, the squeezing is preserved for a time determined by the spins decay rate $\gamma$. In order to better analyze the squeezing dynamics we derived the equation of motion for the squeezing parameter, in the mean-field approximation
\begin{equation}
\label{xit}
\frac{d \xi^2}{dt}
=
-  \big ( i 4 g_{\rm eff} N  \langle\hat a \rangle   + \gamma \big ) \xi^2  + \gamma\, ,
\end{equation}
focusing on its short-time behaviour. In \appref{app:bosonic}, both the bosonic replacement and the mean field approximation have been tested, and the long-time squeezing dynamics is also described.
This squeezing evolution is akin to the one arising from the two-axis twisting Hamiltonian,
which has been shown to be optimal \cite{Ma2011}, so that it determines  $\xi^2 \sim 1/N$ (in absence of decoherence). 

Moreover, it squeezes exponentially in time \cite{Opatrn2015, Liu2011}.
In previous approaches, based on adiabatic field elimination in the bad-cavity limit, the resulting collective spin decay (induced by the cavity) reduces the degree of squeezing. As a consequence, the resulting spin squeezing scales as   $\xi^2 \sim 1/\sqrt{N}$ \cite{Ma2011,Srensen2002, SchleierSmith2010, DallaTorre2013, Borregaard2017}. In our proposal,  the time evolution of the cavity field is taken into account since real photons are involved, and the cavity field, in \eqref{xit}, may add extra dissipation.  
However, its effect is negligible if the spins are able to reach the maximum squeezing faster than the cavity dissipation time scale $\kappa^{-1}$.

\begin{figure}[!ht]
	\centering
	\includegraphics[width=1.02\columnwidth]{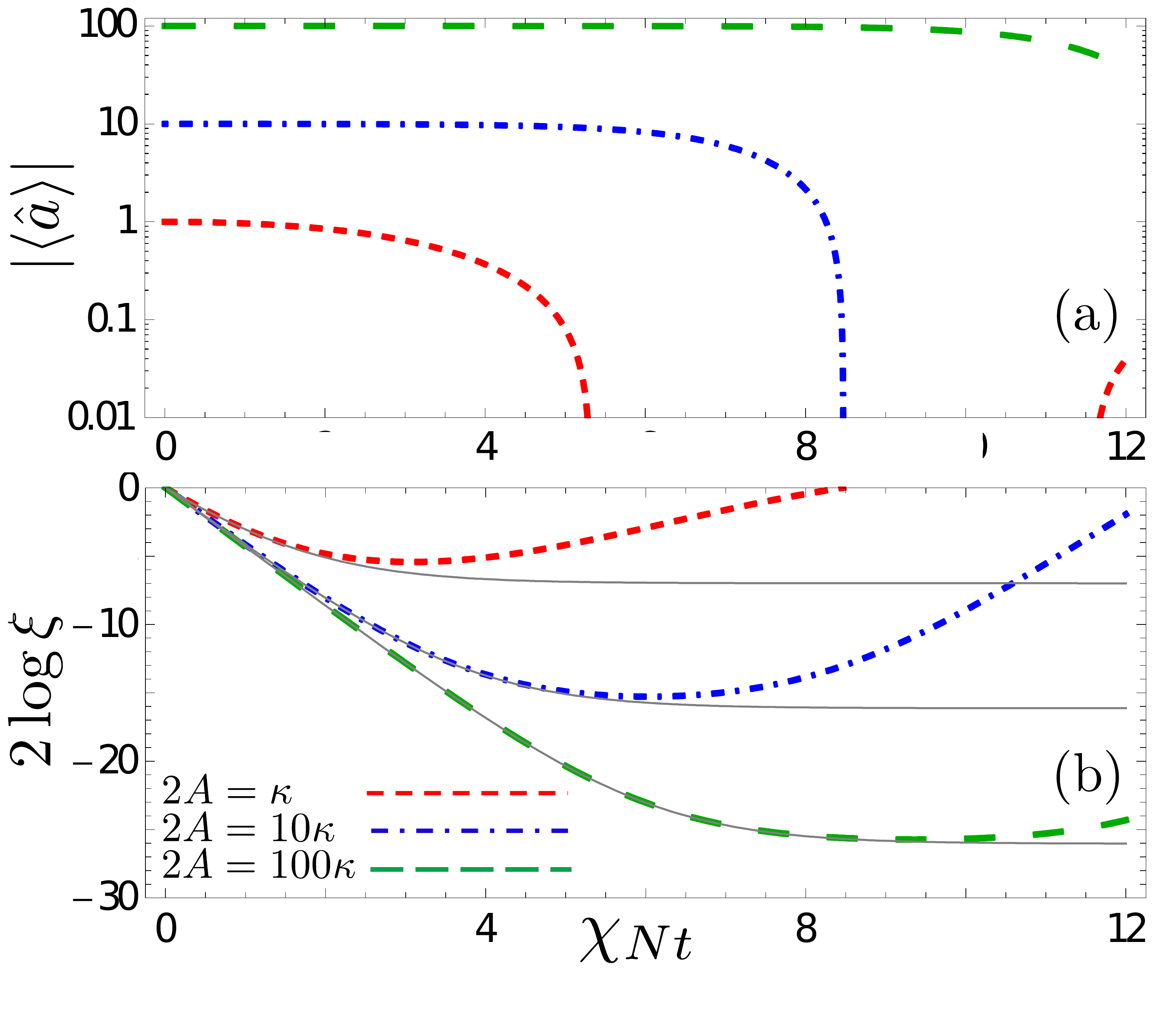}
	\caption{\footnotesize  (a) Absolute value of the cavity field coherence $|\langle \hat a\rangle|$ versus time , and (b) evolution in dB for the squeezing parameter $\xi^2$. Three different values of the drive intensity have been considered: $ 2 A = \kappa, 10 \kappa, 100 \kappa$ (reported in the legend). The rest of the parameters are $N g_{\rm eff} = \kappa = \gamma =1$. The time scale, $(\chi N) ^{-1}$ is given by \eqref{tau}. The solid grey curves describe the results obtained using the analytical result in \eqref{xit2}.}
	\label{fig:boson}
\end{figure}

We propose  a two-step protocol. Owing to the resonant nature of the squeezing mechanism studied here, we start setting the spins out of resonance ($\omega_c  \neq 2 \sqrt{\Delta^2+\epsilon^2}$) and in their ground state. Then, we drive the cavity by a resonant coherent field, until it reaches  $|\langle \hat a \rangle | = \sqrt{n_{\rm ph}}$, where $n_{\rm ph}$ is the steady-state mean photon number in the coherently driven cavity.  Once the cavity is fed, the second step starts: the qubits frequency are tuned  non-adiabatically  into resonance with the cavity.
Numerically, taking as initial condition the spins in their ground state and the cavity in a coherent state with $|\langle \hat a \rangle | = \sqrt{n_{\rm ph}}$, we  compute the squeezing dynamics within the bosonic replacement, as well as the dynamics of the coherent cavity field. Further details are given  in \appref{app:bosonic}.
In \figref{fig:boson} we plot our results. They show  that the maximum  squeezing is obtained in a time scale 
\begin{equation}
\label{tau}
(4 g_{\rm eff} N  \sqrt{n_{\rm ph}} )^{-1}\equiv 
(\chi N)^{-1} \, .
\end{equation}
Notice that the non-adiabatic tuning of the qubits at the beginning of the second step has to occur within a time much lower than the time-scale in \eqref{tau}.

The time at which $\xi^2$ is built up (which marks the short time scale compared to $\kappa$ and $\gamma$ within our parameter regime), can be approximately calculated setting $|\langle \hat a \rangle| =\sqrt{n_{\rm ph}} $ (i.e., constant) in  \eqref{xit}. Then, the dynamics can be solved analytically yielding  the squeezing
\begin{equation}
\label{xit2}
\xi^2 =  \frac{ \chi N\,{\rm exp}[- (\chi  N + \gamma ) t] + \gamma }{\chi N  + \gamma}  \;.
\end{equation}
In \figref{fig:boson}(b) we show that this simple formula explains the attainable squeezing (grey solid curves).
Figure~\ref{fig:boson}(a) displays the time evolution of the absolute value of the mean cavity amplitude $|\langle \hat a \rangle|$ after the  second step of the  protocol. 
At time $t=0$, the cavity field starts from its steady state and the qubits frequency is tuned  non-adiabatically  into resonance with the cavity (second step). As time goes on, the cavity starts to transfer its excitation to the spin ensemble and, as a consequence, $|\langle \hat a(t) \rangle|$ decreases.

This approximation works well if $\chi N \gg \kappa$ (see also figure \ref{fig:boson}(a) and Appendix \ref{app:bosonic}). Moreover, the maximum squeezing is obtained if $\chi N \gg \gamma$ is also satisfied.
These inequalities are largely satisfied for $\kappa \sim \gamma \sim N g_{\rm eff}$.
%
%
Consequently, \eqref{xit2} shows that
$\xi^2$ is reached exponentially and scales  as  $1/N$ (optimal spin squeezing).
However, this result has been obtained after a number of approximations. Although it provides indications that  in the high excitation regime, a significant amount of spin-squeezing can be obtained, it has to be taken with care. Specifically, starting from \eqref{H}, we derived the effective interaction in \eqref{Heff}. Then, we applied to the latter the bosonic approximation. Finally, we employed the mean-field approximation. In addition, we assumed  that the atomic dissipation occurs only through collective interaction with the environment (see Appendices \ref{app:dissipation} and \ref{app:bosonic}). Further studies, including large-scale numerical calculations based on the full quantum model in \eqref{H}, are desirable to confirm the results shown in \figref{fig:boson}.


%

\section{Implementation}
\eqref{xit2} establishes that $\xi^2$ decays exponentially to  $\gamma / (\chi N)$. This term can be rewritten in terms of both, single qubit-cavity coupling and average number of photons in a  driven cavity ($n_{\rm ph}$) such that

\be
 \frac{\gamma }{\chi N}  \sim  \frac{\gamma }{ g N \sqrt{n_{\rm ph}} (g / \omega_q)^2  }. 
\ee
Notice that $\sqrt{n_{\rm ph}}= 2A/\kappa$ ,where $A$ is the drive amplitude.
As said, in the \emph{pros} side is the $N^{-1}$ dependence, together with the enhancement due to the initial driving.  However, in the \emph{cons} we have the ratio $(g/\omega_q)^2$.  
This trade off is better understood by comparing our protocol with similar ones.  
We consider a recent and optimized protocol \cite{LewisSwan2018}, which is limited by spins dissipation as
\be
\xi^2_{\rm Ref. [13]} \sim   \frac{\sqrt{\kappa \gamma/ N } }{g}\,, 
\ee
so that our scaling in terms of the latter will be
\be
\xi^2 \sim \left (\frac{\omega_q}{g} \right)^2  \sqrt{\frac{\gamma }{\kappa  N n_{\rm ph}}} \xi^2_{\rm Ref. [13]}. 
\ee
Hence, we need to search for architectures where: (i) it is possible to couple a large number of effective spins to a single-mode cavity, (ii) together with a non-negligible single-spin normalized  coupling strength ($g/\omega_q$) and (iii) strong cavity pumping is feasible.

Several cavity-QED architectures can satisfy the above criteria. Here, we discuss the coupling between single molecular magnets and coplanar waveguide cavities. Using nano-constrictions together with magnetic molecules, single spin-cavity coupling strengths of the order of $g/\omega_q \sim (10^{-4}-10^{-3})$ can be achieved both for $S=1/2$ and higher spin molecules having decoherence times up to ms \cite{Jenkins2013, jenkins2016}.  
Moreover, since the molecules are nano-sized objects, a macroscopic number of them can be coupled to the coplanar waveguide cavity. In addition, the protocol assumes  that  turning on- and off-resonance the spins indus the fastest time scale in the problem. %
Following \cite{jenkins2016}, in our simulations we consider  realistic numbers $g/\omega_q \sim 10^{-4}$ and $\gamma \sim \kappa \sim 10^{-5}$.  Then, placing  $N \sim 10^9$ of those spins we get the parameters used in Fig. \ref{fig:boson}.
Finally, we must consider the maximum power admitted  in the nanoconstrictions.
Experiments \cite{Jenkins2014} showed that $P_{\rm in} = -40$ dB (which corresponds to a  number of photons $n_{\rm ph}= 10^{12}$) can be safely used. 
Setting these realistic parameters we can get  $g_{\rm eff} N = \gamma = \kappa$  obtaining   $\xi^2 = -30$ dB as shown in Fig. \ref{fig:boson}. These numbers are obtained within our effective theory,  which has also been verified for tens of qubits, although, further studies including large-scale numerical calculations based on the full quantum are needed.

\section{Conclusions}
In this letter we have generalized  the \emph{one photon-two atom process} in cavity-QED to many atoms.  In doing so, we have introduced a {\em novel way for generating many-body spin-spin interactions}.
This yields a two-axis twisting-like interaction among the spins.
Note that the mechanism is a resonant process involving real photons which facilitates the control of the effective interaction. 
We have shown that, already at the single-photon limit, a sizeable squeezing is produced.

Moreover, we showed that by strongly driving  the  cavity, the squeezing scales as $\sim 1/N$, corresponding to the Heisenberg limit. We calculated that reaching $ - 30$ dB is already possible with the spin-cavity and decoherence rates reported for magnetic molecules coupled to superconducting circuit resonators. However, this result is obtained after a number of approximations and thus requires confirmation by further analysis.

Our results could be implemented in various platforms, including several cavity-QED systems; although the on-chip device analyzed here can be particularly advantageous for practical applications.
Here, we focused on generating squeezing, however, the novel spin-spin interaction found here can expand the possibilities for exploring  many-body and nonlinear quantum physics \cite{Stassi2017a}.



\section*{Acknowledgements}
DZ acknowledges RIKEN for its hospitality, the support by the Spanish Ministerio de Ciencia, Innovaci\`{o}n y Universidades within project MAT2017-88358-C3-1-R. The Arag\`{o}n Government project Q-MAD,  EU-QUANTERA project SUMO and the Fundaci\`{o}n BBVA. F.N. is supported in part by the: MURI Center for Dynamic Magneto-Optics via the Air Force Office of Scientific Research (AFOSR) (FA9550-14-1-0040), Army Research Office (ARO) (Grant No. Grant No. W911NF-18-1-0358), Japan Science and Technology Agency (JST) (via the Q-LEAP program, and the CREST Grant No. JPMJCR1676), Japan Society for the Promotion of Science (JSPS) (JSPS-RFBR Grant No. 17-52-50023, and JSPS-FWO Grant No. VS.059.18N), the RIKEN-AIST Challenge Research Fund, the Foundational Questions Institute (FQXi), and the NTT PHI Laboratory. S.S. acknowledges the Army Research Office (ARO) (Grant No. W911NF-19-1-0065).

\newpage

\appendix

\section{Derivation of the effective Hamiltonian: Two-level atoms case}
\label{app:effective}

In order to derive the effective Hamiltonian in \eqref{Heff} (see the main text), we start from \eqref{H}. We first rewrite it in the basis where the qubits Hamiltonian (in the presence of interaction) is diagonal. We obtain
\be \label{V}
\hat H= \omega_q \hat J_z + \omega_c \hat a^\dag \hat a + 2 g (\hat a + \hat a^\dag) (\cos \theta \hat J_x +\sin \theta \hat J_z)\, ,
\ee
where $2 \hat J_\alpha =\sum_i \hat \sigma_\alpha^{i}  $ ($\alpha=x,y,z$).
Notice that the flux offset, is now encoded in the angle $\theta = \arctan (\epsilon / \Delta)$. System Hamiltonian \eqref{V} can reads as sum of two elements: a non-interacting part $\hat H_0= \omega_q \hat J_z + \omega_c \hat a^\dag \hat a$ which describes the bare energy of the system and the light-matter interaction potential part $\hat H_I=2 g (\hat a + \hat a^\dag) (\cos \theta \hat J_x + \sin \theta \hat J_z)$. 

We now apply the generalized James'effective Hamiltonian method \cite{Shao2017} which at the 
third order, neglecting the time dependent terms (RWA), gives the effective interaction Hamiltonian [Eq.~(15) of Ref.~\cite{Shao2017}]
\begin{widetext}
\be \label{veff}
\begin{split}
	\hat H_{\rm eff} =&- \left[\hat h_1 \hat h_2^{\dag} \hat h_1+\hat h_1^{\dag} \hat h_2 \hat h_1^{\dag} \right]+\frac{1}{2}
	\left[\hat h_1 \hat h_1 \hat h_2^{\dag}+\hat h_1^{\dag} \hat h_1^{\dag} \hat h_2+\hat h_2^{\dag} \hat h_1 \hat h_1+ \hat h_2 \hat h_1^{\dag} \hat h_1^{\dag} \right]-\frac{1}{2}
	\left[\hat h_1 \hat h_3^{\dag} \hat h_2+\hat h_2 \hat h_3^{\dag} \hat h_1+\hat h_2^{\dag} \hat h_3 \hat h_1^{\dag}+\hat h_1^{\dag} \hat h_3 \hat h_2^{\dag} \right] \\
	&+\frac{1}{3}
	\left[\hat h_1 \hat h_2 \hat h_3^{\dag}+\hat h_1^{\dag} \hat h_2^{\dag} \hat h_3+\hat h_3^{\dag} \hat h_2 \hat h_1+\hat h_3 \hat h_2^{\dag} \hat h_1^{\dag} \right]+\frac{1}{6}
	\left[\hat h_2 \hat h_1 \hat h_3^{\dag}+\hat h_2^{\dag} \hat h_1^{\dag} \hat h_3+\hat h_3^{\dag} \hat h_1 \hat h_2+\hat h_3 \hat h_1^{\dag} \hat h_2^{\dag} \right]\;,
\end{split}
\ee
\end{widetext}
where $\hat h_1= g \cos{\theta}\, \hat a^{\dag} \sum_{i} \hat \sigma_{-}^{i}$, $\hat h_2=g \sin \theta\,  \hat a^{\dag} \sum_{i} \hat \sigma_{z}^{j}$ and $\hat h_3=g \cos \theta\,  \hat a^{\dag} \sum_{i} \hat \sigma_{+}^{i}$.
Replacing  $\hat h_i$ into \eqref{veff}, adopting normal ordering for the photonic operators and neglecting higher-order terms involving two destruction or creation photon operators, we obtain
\begin{widetext}
\be \label{16}
\begin{split}
		\hat H_{\rm eff} =&-\frac{2g^3 \cos^2 \theta \sin \theta}{\omega^2_{\rm q}} \left[\hat a\sum_{jk} \hat \sigma_{+}^{j}  \hat \sigma_{+}^{k}+\hat a^{\dag}\sum_{jk} \hat \sigma_{-}^{j} \hat \sigma_{-}^{k} \right] \\
		&-\frac{g^3 \cos^2 \theta \sin \theta}{2\omega^2_{\rm q}}
		\left[2\hat a\sum_{ijk} \hat \sigma_{+}^{j}  \hat \sigma_{+}^{k} \hat \sigma_{z}^{i}+4\hat a\sum_{jk} \hat \sigma_{+}^{j}  \hat \sigma_{+}^{k}+2\hat a\sum_{ijk} \hat \sigma_{-}^{j}  \hat \sigma_{-}^{k} \hat \sigma_{z}^{i}-4\hat a\sum_{jk} \hat \sigma_{-}^{j}  \hat \sigma_{-}^{k}\right] \\
		&+\frac{2g^3 \cos^2 \theta \sin \theta}{3\omega^2_{\rm q}}
		\left[\hat a\sum_{ijk} \hat \sigma_{+}^{j}  \hat \sigma_{+}^{k} \hat \sigma_{z}^{i}+2\hat a\sum_{jk} \hat \sigma_{+}^{j}  \hat \sigma_{+}^{k}+\hat a\sum_{ijk} \hat \sigma_{-}^{j}  \hat \sigma_{-}^{k} \hat  \sigma_{z}^{i}-2\hat a\sum_{jk} \hat \sigma_{-}^{j}  \hat \sigma_{-}^{k}\right]\\
		&+\frac{g^3 \cos^2 \theta \sin \theta}{3\omega^2_{\rm q}}
		\left[\hat a\sum_{ijk} \hat \sigma_{+}^{j}  \hat \sigma_{+}^{k} \hat \sigma_{z}^{i}+4\hat a\sum_{jk} \hat \sigma_{+}^{j}  \hat \sigma_{+}^{k}+\hat a\sum_{ijk} \hat \sigma_{-}^{j}  \hat \sigma_{-}^{k} \hat  \sigma_{z}^{i}\right]\,.
	\end{split}
	\ee
\end{widetext}
Given \eqref{16}, after some algebra, we obtain the effective interaction Hamiltonian in terms of the collective lowering and raising spin operators $\hat J_\pm = \sum_ i \hat \sigma^i_\pm$  
\be \label{Heff2} 
\hat H_{\rm eff} =
-\frac{4g^3 \cos^2 \theta \sin \theta}{3 \omega^2_{\rm q}}
\left(\hat a\hat J_{+}^{2}+\hat a^{\dag}\hat J_{-}^{2}\right)\, .
\ee
We note that the the resulting effective interaction Hamiltonian \eqref{Heff2} does not depend on the Pauli operators $\hat J_z$. 

Equation (\ref{Heff2}) displays the effective interaction Hamiltonian, describing the simultaneous generation of two excitations in an ensamble constituted by an arbitrary number $N$ of identical atoms, by one photon absorption. The effective interaction Hamiltonian \eqref{Heff2} is responsible for the coupling between the eigenvectors $ |\textcolor{magenta}{0},ggg..\textcolor{magenta}{e}..\textcolor{magenta}{e}..ggg..\rangle  $ and  $ |\textcolor{magenta}{1},ggg..\textcolor{magenta}{g}..\textcolor{magenta}{g}..ggg..\rangle  $. In terms of the angular momentum notation, they can be written as $ |0,j,-j+2\rangle  $ and $ |1,j,-j\rangle  $, respectively. More generally, this Hamiltonian couples states differing by two-qubit excitations:
$|j,m\rangle \leftrightarrow |j,m+2\rangle$.
The effective resonant coupling between these eigenstates, for $m =-j$ is
\be \label{geffA}
\langle 0, j, -j+2 | H_{\rm eff} | 1, j, -j\rangle=
\frac{4g^3 \cos^2 \theta \sin \theta}{3 \omega^2_{\rm q}}
\sqrt{2N(N-1)}\,.
\ee
\subsection{Derivation of the effective Hamiltonian: $\Delta$-like three-level atoms case} 
It has been shown \cite{Zhao2017} that it is possible to simultaneously excite two atoms by using a cavity-assisted Raman process in combination with a cavity-photon-mediated interaction. We generalize this analysis to a system of many atoms.
Specifically, we consider a system of $N$ $\Delta$-like three-level atoms interacting with a single mode optical resonator \cite{Liu2005,Zhao2017}. The total Hamiltonian is $\hat H_{\Delta} =  \hat H_c +\hat H_{0} +\hat H_I$, being
\be
\hat H_c = \omega_c \hat a^\dag \hat a\, ,
\ee
the energy of the cavity,
\be
\hat H_{0} = \sum_i \left(
\omega_g \hat \sigma^{(i)}_{gg} + \omega_e \hat \sigma^{(i)}_{ee}+\omega_s \hat \sigma^{(i)}_{ss}
\right)\, ,
\ee
the energy of an ensemble of identical three-level ($g,e,s$) atoms  (here, $\hat \sigma^{i}_{mn} = | m \rangle_i {}_i \langle n|$, where $| m \rangle_i$ is a generic eigenstate of the three-level atom with $m,n = g,e,s$), and finally, 
\be\label{V3}
\hat H_I = \hat a \sum^{N}_{i=1} \left(g_{ge} \hat \sigma^{(i)}_{eg} +   g_{gs} \hat \sigma^{(i)}_{sg} + 
g_{es} \hat \sigma^{(i)}_{se} \right) + {\rm H.c.}\, 
\ee 
is the interaction Hamiltonian part. The term $ \hat \sigma^{(i)}_{mn} = \ket{m}\bra{n}$ is a transition operator for the $i$-th three-level atom, while $g_{mn}$ is the corresponding transition matrix elements.
Assuming that the system operates in the dispersive regime: $|\Delta_{mn}| \gg g_{nm}$, where $\Delta_{mn} = \omega_{mn} - \omega_c$ and $\omega_{mn} = \omega_m - \omega_n$ denote the transition frequencies, it is possible to derive the effective Hamiltonian applying a unitary transformation able to eliminate the direct atom-cavity coupling
\be
\hat H_{\rm eff} = e^{- \hat X} \hat H_{\Delta} \, e^{\hat X}\, ,
\ee
where
\be
\hat X =  \sum^N_{i=1} \left(\frac{g_{ge}}{\Delta_{eg}} \hat \sigma^{(i)}_{eg} +
\frac{g_{es}}{\Delta_{se}} \hat \sigma^{(i)}_{se}  + \frac{g_{gs}}{\Delta_{sg}} \hat \sigma^{(i)}_{sg} -{\rm H.c.}\right)\, .
\ee

Keeping terms up to the third order in the interaction Hamiltonian, assuming that no atom is initially in the $|s \rangle$ state, assuming  $\omega_c \simeq 2 \omega_{eg}$, and including only the time-independent terms (in the Heisenberg picture), we obtain the effective Hamiltonian
\be \label{Heff4} 
\hat H_{\rm eff} = g_{\rm eff}
\left(\hat a\hat J_{+}^{2}+\hat a^{\dag}\hat J_{-}^{2}\right)\, ,
\ee
with the resulting effective coupling strength 
\be	
g_{\rm eff} = \frac{g_{ge} g_{gs} g_{se}}
{ 3 \Delta_{ig} \Delta_{ie} {\rm \Delta}_{eg}}
(3 \Delta_{ig}- \omega_{eg}) \, .
\ee	
Notice that in \eqref{Heff4} $\hat J_{+} = \sum_i \hat \sigma^{(i)}_{eg}$.

\section{Comparison of energy levels and system dynamics obtained using the effective and the full models.}
\label{app:Full}

Here we start comparing the lowest energy levels [see \figref{fig:spectrum}], obtained by using the effective Hamiltonian in \eqref{Heff},   with the ones calculated  by using the full system Hamiltonian in \eqref{V} (which is equivalent to \eqref{H}).

\begin{figure*}[!htbp]
	\includegraphics[width=18cm]{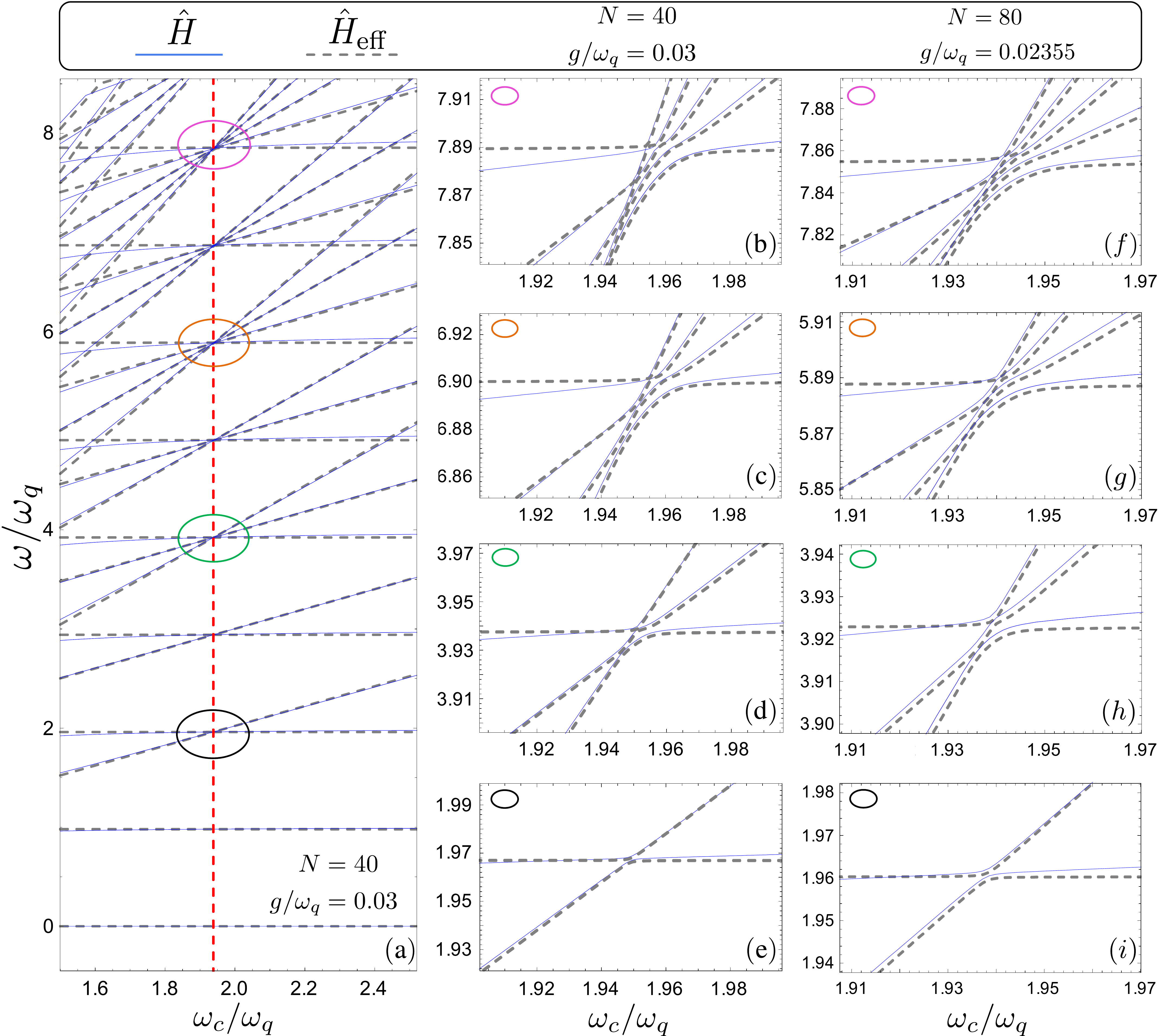}
	\caption{\footnotesize (a) Lowest energy levels of the full (blue solid curves) and the effective Hamiltonian (grey dashed curves) as a function of $\omega_c/ \omega_q$, obtained for $N=40$ qubits and individual coupling strength $g/\omega_q=0.03$.  Panels~\ref{energy_level}(b-e) display an enlarged view of the first four avoided level crossings [indicated by colored circles in panel~(a)]. Panels~\ref{energy_level}(f-i) report the same first four avoided level crossings, obtained  for $N=80$ qubits, decreasing the individual coupling strength ($g/\omega_q=0.02355)$ in order to keep constant the energy splittings.}\label{energy_level}
\end{figure*}

Figure~\ref{energy_level}(a) shows the lowest energy levels of the full Hamiltonian (blue solid curves) and those obtained diagonalizing the effective Hamiltonian (gray dashed curves) as a function of $\omega_c/ \omega_q$, calculated for $N=40$ qubits and using an individual coupling strength $g/\omega_q=0.03$. Since the considered effective model does not include the renormalization of the bare energy levels induced by the light-matter interaction in the dispersive regime, the bare transition energy of the spin has been used as the only fitting parameter.
A ladder of well-aligned avoided level crossings at $\omega_c  \simeq 2 \omega_q$ (highlighted with color circles) are clearly visible. Panels~(b-e) show an enlarged view of the first four avoided level crossings, indicated by circles in panel~(a).

We observe a number of interesting features: (i) the four avoided level crossings in the figure [see also the enlarged views in panels (b-e)] are well aligned [see the vertical dashed red line in (a)], all at the same value of $\omega_c / \omega_q$. This vertical alignment in (a) is also preserved for (not shown) the higher energy avoided crossings for values $\omega_ / \omega_q$ well below $N$; (ii) the agreement between the displayed levels obtained using the full and the effective models is good for values of $\omega_c / \omega_q \simeq 2$ around the minima of the avoided level crossings (resonance condition); (iii) the agreement is less good moving away from the resonance condition. This is simply due to the dependence (not taken into account) of the energy shifts on the bare cavity frequency.
In Figs.~\ref{energy_level}(f-i) we plot the same first four avoided level crossings for $N=80$ qubits, decreasing the individual coupling strength $g/\omega_q=0.02355$ in order to keep constant the energy splittings. Clearly, both the full and effective models  are still in good agreement at $\omega_c  \simeq 2 \omega_q$.

\begin{figure}[!htbp]
	\includegraphics[width=9cm]{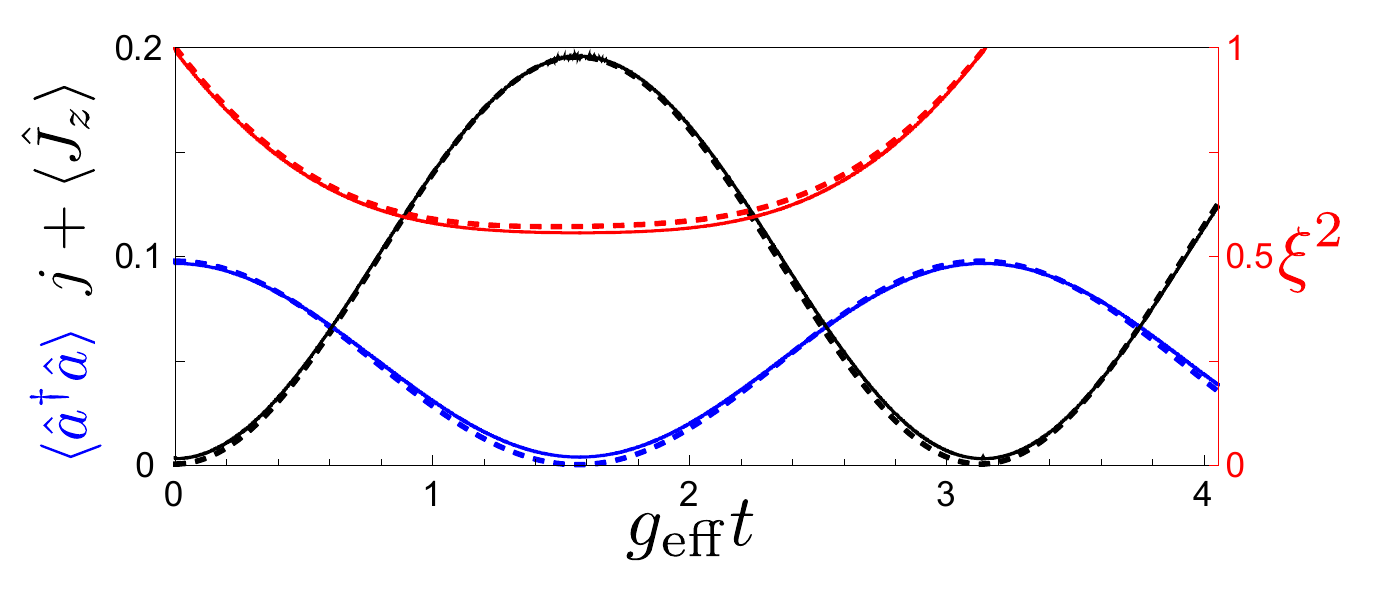}
	\caption{\footnotesize Free time evolution of the interacting light-matter system, considering $N=10$ effective spins  for $ \omega_c= 2\omega_q $. The initial state is a superposition of the system ground state and a one-photon state with all the spins in their ground state (see the main paper). The continuous curves describe the mean number of cavity photons (blue) as well as the mean excitation number for the spin system (black), and the squeezing parameter (red) obtained using \eqref{H} (numerical calculation). The dashed curves correspond to the analytical results displayed in Fig.~\ref{fig:spin}(a) for the effective Hamiltonian \eqref{Heff}.}\label{dynamics}
\end{figure}

As a further check, we now analyze the free system evolution considering, as initial condition, a superposition state of the system ground state and a one-photon state with all the spins in their ground state (see the main paper \secref{single_photon}).
All the  parameters used here coincide with those used to obtain the results in Fig.~\ref{fig:spin}(a). 
Figure~\ref{dynamics} displays such a comparison.  Specifically, the continuous curves describe the mean number of cavity photons (blue) as well as the mean excitation number for the spin system (black), and the squeezing parameter (red) obtained using \eqref{H} (the solid curves show the numerical calculations). The dashed curves in Fig.~\ref{dynamics} correspond to the analytical calculations displayed  in Fig.~\ref{fig:spin}(a).
The agreement between the two sets of curves (dashed and continuous curves in Fig.~\ref{dynamics}) is very good, showing that the effective Hamiltonian is able to describe well this interacting system under the resonant condition $\omega_c \simeq 2 \omega_q$, at least for a moderate light-matter interaction strength. 

\section{Dissipation}
\label{app:dissipation}

The Linblad dissipators for a collection of two level systems inside a cavity have been discussed in Ref.\,\cite{Gelhausen2017}.
In this appendix we develop a similar theory adapting it to our case.
We assume that the ensemble of spins occupies a small volume, compared to the cavity-mode wavelength.
This is the case for the implementation discussed in the main text.
With this assumption, both the cavity and atomic decays can be cast in the 
Lindblad form \cite{Gelhausen2017}
\begin{align}
\label{kDa}
\kappa \mathcal D [\hat a] & = \kappa  \big ( 2 \hat a \hat \varrho \hat a^\dagger - \{ \hat a^\dagger \hat a
, \hat \varrho \} \big )
\\
\label{gDs}
\gamma \mathcal D [\hat \sigma] & =
\gamma  
\sum_i \hat \sigma_-^i \hat \varrho \hat \sigma^i_+ - \frac{1}{2} \{ \hat \sigma_+^i
\hat \sigma_+^i, \hat \varrho \}\,.
\end{align}
Here, $\gamma$ and $\kappa$ are the atomic and cavity decay rates, respectively.
Taking the Fourier transform,
\begin{equation}
\label{s0S1}
\hat \sigma^k_+ = \frac{1}{\sqrt{N}} \sum_j e ^{i k j} \hat \sigma^j_k 
\end{equation}
we notice that 
\begin{equation}
\label{s0S}
\hat \sigma^0_\pm = \frac{1}{\sqrt{N}} \hat J_\pm \; .
\end{equation}
Using \eqref{s0S1}, the single-site dissipative terms becomes:
\begin{align}
\label{transmom}
&\sum_i \hat \sigma_-^i \hat \varrho \hat \sigma^i_+ - \frac{1}{2} \{ \hat \sigma_+^i
\hat \sigma_+^i, \hat \varrho \}\\ \nonumber
&=\frac{1}{N}
\sum_{k, k^\prime} \sum_j   e^{i ( k - k^\prime) j} 
\Big (
\hat \sigma_-^k \hat \varrho \hat \sigma^{k^\prime}_+ - \frac{1}{2} \{ \hat \sigma_+^{k^{\prime}}
\hat \sigma_-^k, \hat \varrho \}
\Big )
\; .
\end{align}
It is convenient to
separate the zero momentum contribution which,  using \eqref{s0S}, results into:
\begin{align}
\label{Dsigma}
\gamma \mathcal D [\hat \sigma]
= &
\frac{\gamma}{ N}  
\big ( 
\hat J_- \hat \varrho \hat J_+ - \frac{1}{2} \{ \hat J_+ \hat J_-, \hat \varrho \} 
\big ) 
\\  \nonumber
& + 
\gamma 
\sum_{k\neq 0}  \hat \sigma_-^k \hat \varrho \hat \sigma^{k}_+ - \frac{1}{2} \{ \hat \sigma_+^{k}
\hat \sigma_+^k, \hat \varrho \}
\; .
\end{align}
Finally, we analize how the terms in the Hamiltonian (\ref{H})  looks like in momentum space. For that, we realize that [Cf. \eqref{transmom}]
\begin{equation}
\hat \sum_i \hat \sigma^i_+ \hat \sigma^i_- = \hat \sum_k \hat \sigma^k_+ \hat \sigma^k_- \; .
\end{equation}
Morover, the atomic-light coupling becomes
\begin{align}
g (\hat a + \hat a^\dagger) \sum_i \hat \sigma_x^i
= g  (\hat a + \hat a^\dagger) ( \hat J_+ + \hat J_-)\,.
\end{align}
As expected, the cavity only couples to the zero momentum operator. Therefore,  the full dynamics (unitary + dissipative) do not mix different momenta, resulting in the QME used in the main text.

\section{ Bosonic map in the large-$N$ limit}
\label{app:bosonic}
If $N$ is sufficiently large and the number of spin excitations satisfies the condition $\sum_j \langle \hat \sigma^+_j \hat \sigma_j^- \rangle \ll N$, the collective spin operator can be replaced \cite{Hummer2012} by a bosonic mode
	\begin{align}
    \label{bos-map}
    \hat J_- \equiv  \sum_i \hat \sigma^{(i)}_- \to  \sqrt{N}\,\hat b \;\;\;\\ \label{bos-map2}
    	\hat J_+ \equiv  \sum_i \hat \sigma^{(i)}_+ \to  \sqrt{N} \, \hat b^\dag\, ,
	\end{align}
where $\hat b $ ($\hat b^\dag $) is the creation (destruction) operator in the new bosonic rappresentation. Using \eqref{bos-map}, \eqref{bos-map2} and including a continuum driving term $ {2 A} \cos (\omega_d \, t)  (\hat a + \hat a^\dagger)$ (in the rotating wave approximation), the system Hamiltonian in the rotating frame becomes:
\begin{equation}
\label{Hboson}
\hat H= (\omega_c-\omega_d)\hat a^\dag \hat a + \omega_q \hat b^\dag \hat b +g_{\rm eff} N \; (\hat a \hat b^{\dagger \; 2} + \hat a^\dagger \hat b^{2}) + A (\hat a +\hat a^\dagger).
\end{equation}
Notice that, since $(\hat b^\dagger)^2 |0 \rangle = \sqrt{2}|2\rangle$, the anticrossing scales as $\sqrt{2}\,N$ which equals $\sqrt{2 N(N-1)}$, in the $N\to \infty$ limit [Cf.  \eqref{geff} in the main text].

As regards the dissipators, they are global spin operators (see \appref{app:dissipation}). Thus, after the replacement by a bosonic mode the  master equation becomes
\begin{align}
\label{qme-boson}
\dot {\hat \varrho} =& -i [\hat H_{\rm bosonic}, \hat \varrho ] 
\\ \nonumber
& + \kappa  \big ( 2 \hat a \hat \varrho \hat a^\dagger - \{ \hat a^\dagger \hat a
, \hat \varrho \} \big ) 
+
\gamma 
\big ( 
\hat b \hat \varrho \hat b^\dagger - \frac{1}{2} \{ \hat b^\dagger \hat b, \hat \varrho \} \big )\,.
\end{align}
Due to the nonlinearity of the system Hamiltonian,  \eqref{qme-boson} is not exactly solvable.  
Applying the mean-field approximation $\langle \hat a \hat b^\dagger\rangle \to \langle \hat a \rangle \langle \hat b^\dagger \rangle$, with the purpose to describe the cavity-spins interaction, we end up with a non-linear and closed  set of coupled equations for the first and second moments
\begin{subequations}
	\begin{align}
	\label{<a>}
	\partial_t \langle \hat a \rangle &= - i N g_{\rm eff} \langle\hat  b^2 \rangle
	- i A - \frac{\kappa}{2} \langle \hat a \rangle
	\\ \label{<b>}
	\partial_t \langle \hat b \rangle &= - i N g_{\rm eff}  \langle \hat a \rangle \langle \hat b^\dagger \rangle
	- \frac{\gamma}{2} \langle \hat b \rangle
	\\
	\label{b2}
	\partial_t \langle \hat b^2 \rangle &= - i 2 N g_{\rm eff} \langle \hat a \rangle ( 2 \langle \hat b^\dagger \hat b \rangle + 1)
	- \gamma \langle \hat b^2 \rangle
	\\ \label{bdb}
	\partial_t \langle \hat b^\dagger \hat b \rangle &= - i 2 N g_{\rm eff}( \langle \hat a \rangle  \langle \hat b^{\dagger \;2} \rangle - {\rm c.c.}) - \gamma \langle \hat b^\dagger \hat b \rangle \; .
	\end{align}
\end{subequations}

\begin{figure}[!t]
	\centering
	\includegraphics[width=1\columnwidth]{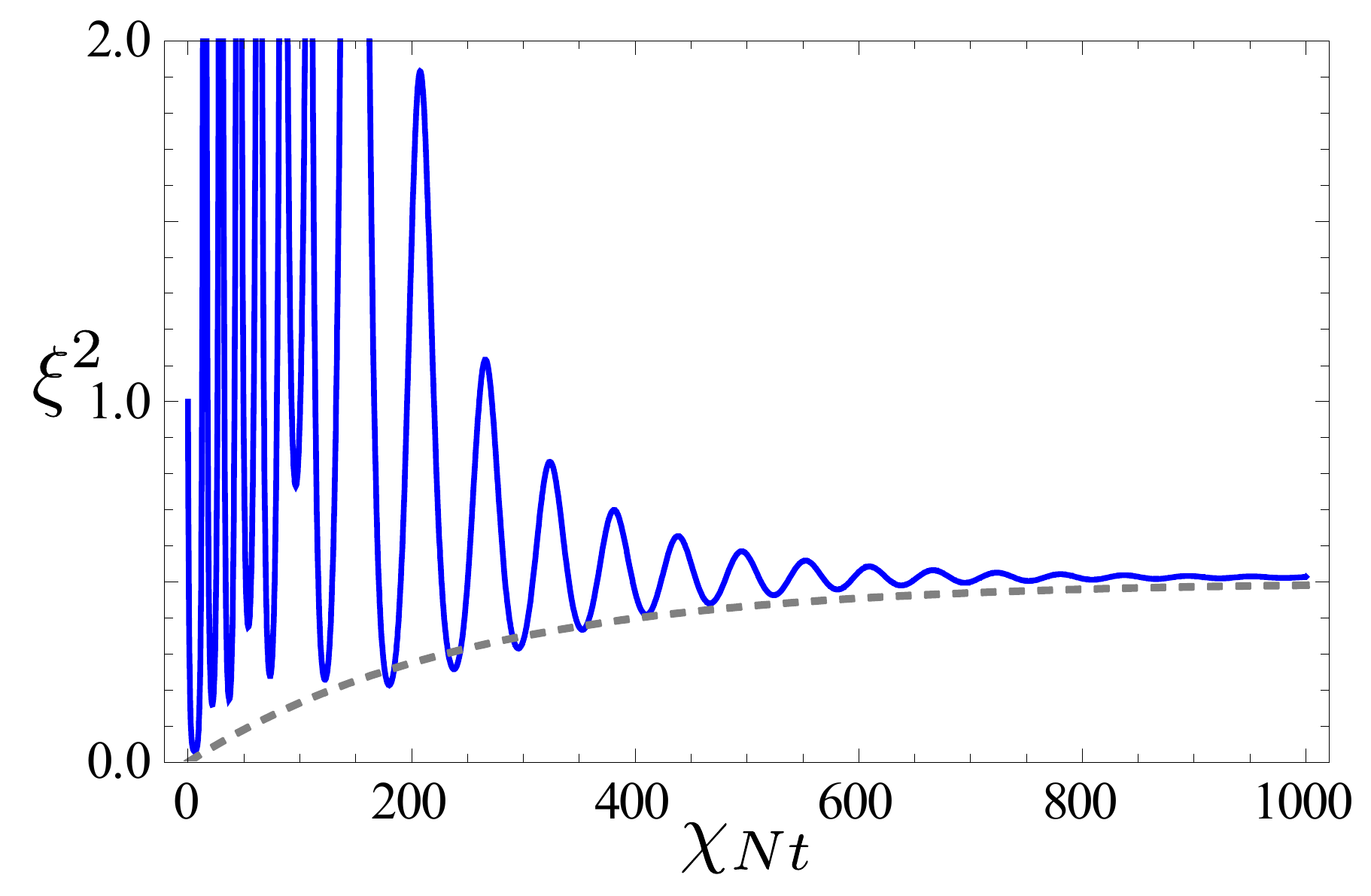}
	\caption{\footnotesize Dynamics for $\xi^2$ using \eqref{xit-app}.  The parameters used are $A= 10\, \kappa$ and $N g_{\rm eff} = \kappa = \gamma =1$}
	\label{fig:fulldyn}
\end{figure}

\begin{figure}[!t]
	\centering
	\includegraphics[width=1.05\columnwidth]{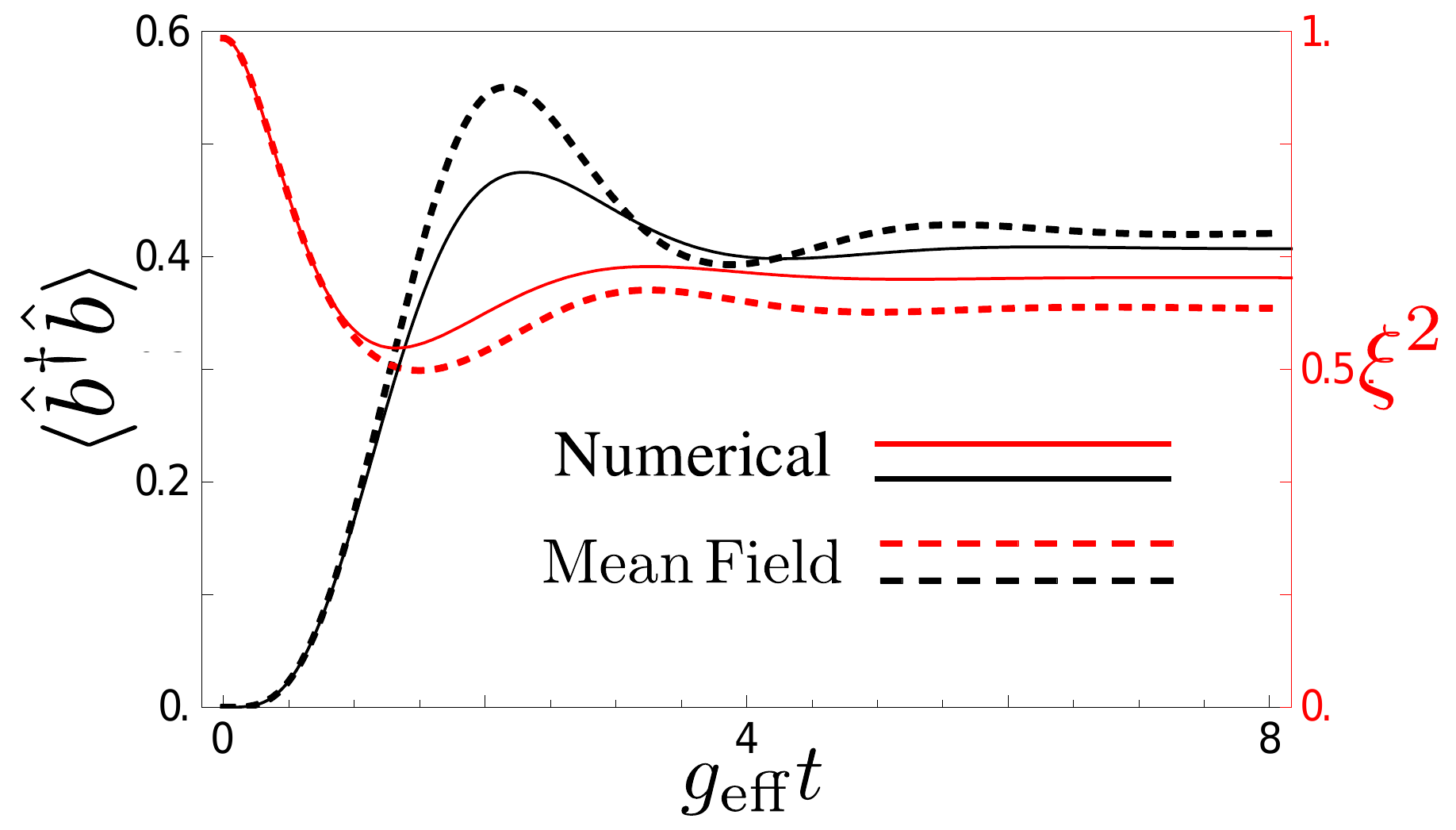}
	\caption{\footnotesize Comparison between the system dynamics solved by numerical calculations and using the mean-field approximation. We compare both the mean number $\langle \hat b^\dagger \hat b \rangle$ and $\xi^2$. The parameters used are $\gamma =5 \times 10^{-2}\,\omega_q $, $A=  \gamma$, $\kappa = \gamma$ and $N g_{\rm eff} = \kappa = \gamma =1$.
	}
	\label{fig:comparison}
\end{figure}

Here, we are interested in computing $\xi^2$, wich in the bosonic limit reads
\cite{Wang2003} 
\begin{equation}
\label{sboson-app}
\xi_{N \to \infty}^2= 1 + 2 \Big ( \langle \hat b^\dagger \hat b \rangle - |\langle \hat b^2 \rangle | \Big ) \; .
\end{equation}  
Besides, to findto find a closed equation of motion for $\xi$ we realize that, at the steady state, in the regime where $g_{\rm eff} \ll 1$, \eqref{<a>} yields that the mean value of the cavity field coherence is a purely imaginary number, $\langle \hat a \rangle= -i 2A/\kappa $. As a consequence, using  \eqref{b2} and \eqref{bdb}, the mean value of the quadratic bosonic operator $\langle \hat b^2 \rangle $ is a  real number. 
Thus, we can replace $|\langle \hat b^2 \rangle|\rightarrow \langle \hat b^2 \rangle $ in \eqref{sboson-app} and, taking the time derivative of it we get
\be
\begin{split}
\frac{d \xi^2_{N \to \infty}}{d t}&= 4 \gamma \left [ \langle \hat b^\dagger \hat b \rangle + \langle \hat b^2 \rangle \right ]\\
&-  i  4 N g_{\rm eff} \langle \hat a \rangle \left [ 1 + 2 ( \langle \hat b^\dagger \hat b \rangle + \langle \hat b^2 \rangle  \right ] \; ,
\end{split}
\ee
ending up to \eqref{xit} (see the main text), namely
\begin{equation}
\label{xit-app}
\frac{d \xi^2}{dt}
= -  \big ( i 4 g_{\rm eff} N  \langle \hat a \rangle   + \gamma \big ) \xi^2  + \gamma\, .
\end{equation}
In \figref{fig:fulldyn}, we plot an example for the time dynamics of $\xi^2$.  Finally,  in \figref{fig:comparison} we test the mean-field approximation comparing the system dynamics solved by numerical and analytical (using mean-field approximation) calculations. We compare both the mean number $\langle \hat b^\dagger \hat b \rangle$ and $\xi^2$, finding a good agreement.

Finally, for completeness, 
let us explore the squeezing obtained in the limit $t\to \infty$ (stationary squeezing) setting the l.h.s of \eqref{<a>}, \eqref{b2} and \eqref{bdb} to zero. First, we introduce some dimensionless quantities, namely
\begin{equation}
\Ak := 2 {\mathcal A} / \kappa \, ,
\quad
\Gg := 2 N g_{\rm eff} / \gamma \, ,
\quad
\Gk := 2 N g_{\rm eff} / \kappa \, .
\end{equation}
From \eqref{b2} and \eqref{bdb}, we solve for $\langle \hat b^\dagger \hat b \rangle$ in the stationary state, obtaining
\begin{equation}
\label{bdbsol}
\langle \hat b^\dagger \hat b \rangle =
\frac{2 \Gg^2 |\langle \hat a \rangle |^2}{1- 4 \Gg^2 |\langle \hat a \rangle |^2} \; ,
\end{equation}
while using \eqref{<a>} and \eqref{b2}  we also get
\begin{equation}
\label{aeq}
\langle \hat a \rangle = - i \Ak + \Gk \Gg \langle \hat a \rangle ( 2 \langle \hat b^\dagger \hat b \rangle + 1)
\; .
\end{equation}
Combining \eqref{bdbsol} with \eqref{aeq} an equation for   $\langle \hat a \rangle$ is obtained
\begin{equation}
\langle \hat a \rangle + \frac{\Gk \Gg \langle \hat a \rangle}{1- 4 \Gg^2 |\langle \hat a \rangle |^2}= - i \Ak\,,
\end{equation}
from which it follows that 
\begin{equation}
\label{xifull}
\xi_{N \to \infty}^2 = 1 + 2 (\langle \hat b^\dagger \hat b \rangle  - |\langle \hat b^2 \rangle|)=
\frac{1}{1+ 2 \Gg |\langle \hat a \rangle|}\,.
\end{equation}     
The minimum of stationary  squeezing is
$  {\rm min} ( \xi^2_{N \to \infty} ) = 1/2$.
This can be understood by looking to \eqref{aeq}.
There, it is easily checked that, when $\Ak \to \infty$, then $1- 4 \Gg^2 |\langle \hat a \rangle |^2 \to 0$, thus $\xi_{N \to \infty}^2 \to 1/2$.
The latter result has been verified numerically.

\newpage

\bibliography{Riken}

\end{document}